# Prediction of Weyl semimetal, AFM topological insulator, nodal line semimetal, and Chern insulator phases in $Bi_2MnSe_4$

Sugata Chowdhury[1*], Kevin F. Garrity[1], Francesca Tavazza[1]

[1]Materials Measurement Laboratory, National Institute for Standards and Technology, Gaithersburg, MD 20899

**Abstract:** Three dimensional materials with strong spin-orbit coupling and magnetic interactions represent an opportunity to realize a variety of rare and potentially useful topological phases. In this work, we use first principles calculations to show that the recently synthesized material $Bi_2MnSe_4$ displays a combination of band inversion and magnetic interactions, leading to several topological phases. $Bi_2PbSe_4$, also studied, also displays band inversion and is a topological insulator. In bulk form, the ferromagnetic phase of $Bi_2MnSe_4$ is either a nodal line or Weyl semimetal, depending on the direction of the spins. When the spins are arranged in a layered antiferromagnetic configuration, the combination of time reversal plus a partial translation is a new symmetry, and the material instead becomes an antiferromagnetic topological insulator. However, the intrinsic TRS breaking at the surface of $Bi_2MnSe_4$ removes the typical Dirac cone feature, allowing the observation of the half-integer quantum anomalous Hall effect (AHC). Furthermore, we show that in thin film form, for some thicknesses, $Bi_2MnSe_4$ becomes a Chern insulator with a band gap of up to 58 meV. This combination of properties in a stoichiometric magnetic material makes $Bi_2MnSe_4$ an excellent candidate for displaying robust topological behavior.

**Introduction**

Since the discovery of time-reversal invariant ($Z_2$) topological insulators[1-4] a decade ago, there has been a major increase in interest in topological condensed matter systems, due to both scientific interest and potential applications.[4-10] Despite these efforts, it remains challenging to find robust materials with realizations of many phases, especially those with broken time reversal symmetry (TRS). For example, Chern insulators,[11] which are two-dimensional insulators that display the quantum anomalous Hall effect, were first demonstrated by Haldane in 1988,[12] but their first material realizations were not created until recently.[13] While current quantum anomalous Hall materials, based on magnetically doped topological insulators,[14-18] are limited to very low temperatures (~1K),[13] there is no intrinsic reason for this limit, and there remains significant interest[19-24] in finding Chern insulators with larger band gaps and higher magnetic ordering temperatures.[24-29]

There are many additional related topological phases with broken TRS that are also difficult to realize experimentally. For example, Weyl semimetals (WSM)[30-32] are materials with topologically protected linearly dispersing band crossings, known as Weyl points, near the Fermi level. WSM require the breaking of at least one among inversion symmetry and TRS.[33-38] Broken inversion WSM have been studied experimentally,[39-47] while TRS-broken WSM have been more difficult to verify.[37, 48-51] They are, however, very interesting as they have different and in some cases simpler properties, such as potentially a minimum two Weyl points in the Brillioun zone, as well as nonzero anomalous Hall conductivity.[34, 35] Furthermore, the interplay between TRS invariant topological insulators and broken TRS materials can lead to a variety of interesting behaviors related to the quantized topological contribution of the magneto-electric effect: $S_\theta = \frac{\theta}{2\pi} \frac{e^2}{h} \int d^3x dt \vec{E}\cdot\vec{B}$.[2, 19, 52, 53] where θ is a dimensionless phase that ranges from 0 to $2\pi$. θ=0 in trivial insulators with TRS, but θ=$\pi$ in $Z_2$ topological insulators with TRS, making $Z_2$ materials potentially useful magnetoelectrics.[54-56] However, to observe this effect, it is necessary to remove the topologically protected metallic surface state of the $Z_2$ material by breaking TRS on the surface.[57-59]

In this work, we investigate the electronic properties of $Bi_2MSe_4$ (BMS, M = Pb, Mn) using density functional theory (DFT) and Wannier-based tight-binding models, and we predict a series of topologically non-trivial phases. Recently, we have reported the growth of a thin film of magnetic systems with stoichiometric composition $Bi_2MnSe_4$ by molecular beam epitaxy (MBE).[60] We demonstrated that the introduction of an elemental beam of Mn during the molecular beam epitaxial growth of $Bi_2Se_3$ results in the formation of layers of $Bi_2MnSe_4$ that intersperse layers of pure $Bi_2Se_3$. Here, we concentrate on the topological properties of this crystal structure. First, we look at $Bi_2PbSe_4$, which we find to be a $Z_2$ topological insulator due to band inversion at the Z point. In the rest of this letter, we focus on combining this band inversion with the magnetic properties of $Bi_2MnSe_4$. We find that depending on the magnetic ordering and thickness, $Bi_2MnSe_4$ can be a nodal line system, magnetic Weyl semimetal, antiferromagnetic topological insulator, or Chern insulator, in addition to displaying the half-integer quantum anomalous Hall effect. In this work we have considered three different materials and two ($Bi_3Se_4$ and $Bi_2MnSe_4$) of them have been synthesized, whereas $Bi_2PbSe_4$ hasn't been synthesized yet. All these materials are stoichiometric and have significant band gaps, which should help make these topological phases robust and possible to observe at increased temperature.

**Crystal Structure**

$Bi_2MSe_4$, like $Bi_2Se_3$, is a layered material, with $R\bar{3}m$ space group (Fig. 1). The $Bi_2MnSe_4$ septuple layer (SL) shown in Fig. S1a, is a 2D layer consisting of seven atomic layers with stacking sequence of Se-Bi-Se-M-Se-Bi-Se along the c-axis (Fig. 1). The individual 2D layers are weakly bound to each other by van der Waals forces. In this work we considered bismuth (Bi), or lead (Pb) or manganese (Mn) as the metal atom (M), which is the inversion center. The optimized lattice parameters of BMS have been tabulated in Table S1.

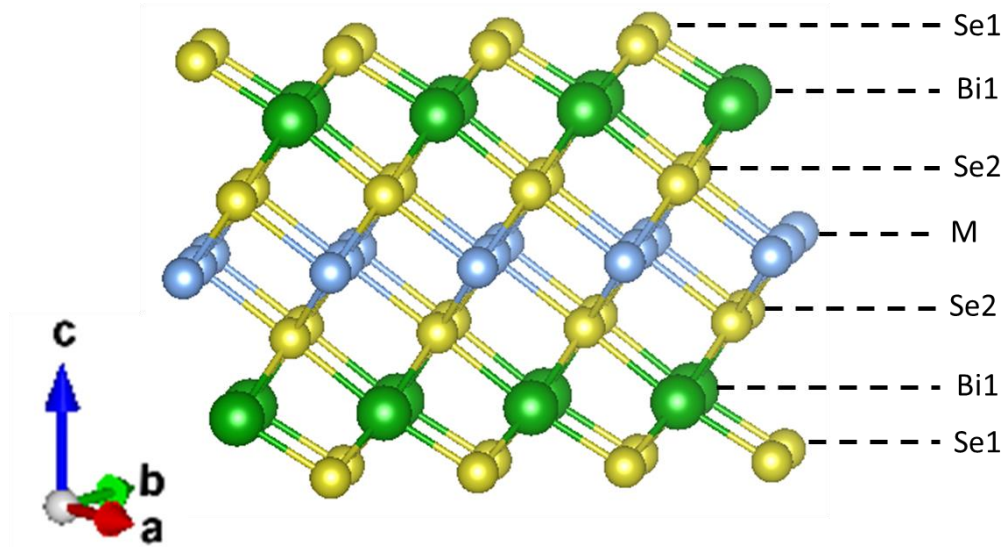

**Fig 1**: Crystal structure of $Bi_2MSe_4$ (M = Bi, Pb, Mn). Here M is the inversion center of the crystal structure, whereas yellow, green and blue sphere are the Se, Bi and M atom respectively.

**Computational**

Calculations were carried out using density-functional theory (DFT)[61, 62] as implemented in QUANTUM ESPRESSO code.[63] We used the PBEsol generalized gradient approximation as exchange and correlation potential.[64] We have used fully relativistic norm-conserving pseudopotentials.[65, 66] We use a plane-wave cutoff of 70 Ry, a 10x10x6 Monkhorst-Pack[67] k-point mesh for bulk geometries and a 8 × 8 × 1 k-point grid for slab geometries. For calculations with Mn, we use DFT+U[68, 69] with U=3 eV, although we find that our results are robust to changes in the value of U, as the states near the Fermi level are not predominantly Mn states. All the geometric structures are fully relaxed until the force on each atom is less than 0.002 eV/Å, and the energy-convergence criterion is 1x10$^{-6}$eV. Results from our DFT calculations are then used as input to construct maximally localized Wannier functions using WANNIER90.[70, 71] For surface state calculation we have considered [001] surface and we divided the semi-infinite system into a surface slab with finite thickness, and a bulk (the remaining part). Topological numbers and band gaps are calculated using Wannier interpolation of band structure using the Wannier-TOOLS package.[72]

**Results**

We begin our discussion of the $Bi_2MSe_4$ structure by considering $Bi_3Se_4$.[73] Each septuplet building block of $Bi_3Se_4$ is comprised of seven atomic (Se-Bi1-Se-Bi2-Se-Bi1-Se) layers along the *c* axis and separated by a van der Waals gap. The optimized lattice constants ($a_{BMS}$ = 4.28 Å and $c_{BMS}$ = 40.9 Å) are in good agreement with experimental results[73]. Relaxed structural parameters, such as bond length and intralayer distance, have been tabulated in Table S1 (Supplementary). As expected by electron counting, our band structure calculations reveal that $Bi_3Se_4$ is a metallic system,[74] as shown in Fig. 1. However, we expect that substituting a (+2) ion for one Bi will exactly fill the Se states, possibly opening a gap.

Accordingly, we replace the Bi2 atom by with Pb, creating $Bi_2PbSe_4$ (BPS). The calculated bulk band structure, shown in Fig. 2a, demonstrates that BPS has a bulk band gap of 0.15 eV. We calculate the topological invariants of BPS, and we find that, like the closely related $Bi_2Se_3$, BPS is a strong ($Z_2$) topological insulator, although the topological band inversion happens at the Z point instead of the Γ point in $Bi_2Se_3$. This is confirmed by the surface local density of states (LDOS) (Fig. 2b), calculated using our Wannier tight-binding parameters, which clearly shows a single Dirac cone at the Γ point in the surface band structure and confirms that BPS is topologically non-trivial.

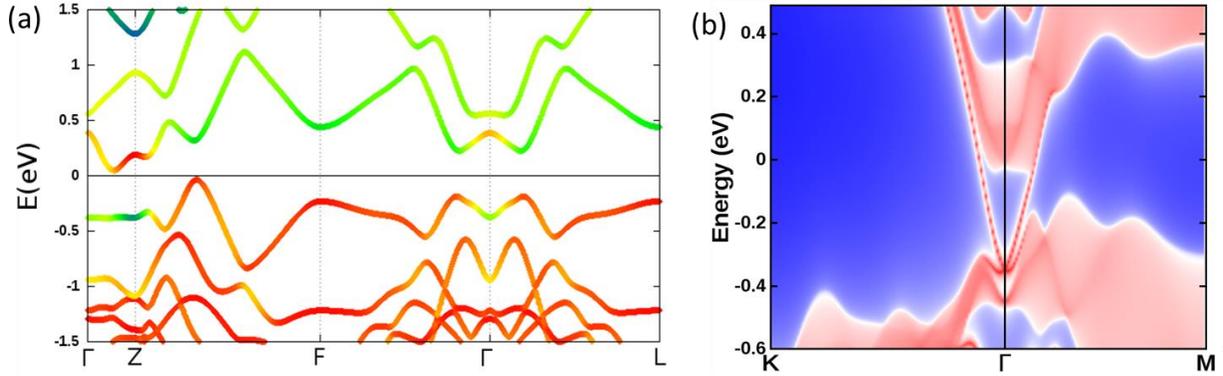

**Fig 2**. (a) The bulk band structure of $Bi_2PbSe_4$ including spin-orbit. (b) Energy and momentum dependence of local density of states (LDOS) of the (001) surface of $Bi_2PbSe_4$. Here, the red region shows the bulk energy bands and the blue region shows the energy gap.

Encouraged by these results, we next investigate the magnetic material $Bi_2MnSe_4$ (BMS), which we will concentrate on for the rest of the paper. Consistently with Otrokov *et. al.,*[75] who studied $Bi_2Se_3$/$Bi_2MnSe_4$ interfaces, we find that the spins prefer to orient ferromagnetically in the case of a single BMS layer. We focus on the topological properties of two spin orderings, ferromagnetic and antiferromagnetic, with alternating layers along the (001) direction (Fig. S3). We find that antiferromagnetic BMS is 1.5 meV per formula unit lower in energy than the ferromagnetic phase, but given the small energy difference it may be possible to align the spins in an external field to observe the ferromagnetic state.

We first consider the ferromagnetic band structure with spins oriented in the (001) direction, as shown in Fig. 3a. By looking at the atomic character of the bands, shown by the colors in Fig. 2a, we can see that there is band in inversion at Γ, with a Bi band well below the Fermi level. We also note that there is spin splitting on the order of 0.1 eV, even though the states near the Fermi level all originate from Bi and Se. At the Fermi level, there are two linearly dispersing band crossing points, at +/- 0.021 Ang.$^{-1}$ along the line from Γ to Z, shown more clearly in Fig. 3a. By calculating the Chern number on a surface surrounding each point, we show that the crossings are topologically protected Weyl points due to band inversion at Γ of bands with $Γ_{4-}$ and $Γ_{5-}$ symmetry. Along the line from Γ to Z, the bands that cross at the Weyl points are characterized by $Γ_4$ and $Γ_5$ symmetry; however, we emphasize that the Weyl points are topologically protected, and breaking symmetries can move the crossings off the high symmetry line, but not gap them. Unlike Weyl semimetals with TRS but broken inversion, which must have at least four Weyl points, $Bi_2MnSe_4$ has the minimum two Weyl points, and they are both exactly at Fermi level due to inversion symmetry[34]. As shown in Fig. S2, we confirm the topological character of the bulk by calculating the surface band structure using our Wannier tight-binding model, finding the expected surface Fermi arcs connecting the two Weyl points.

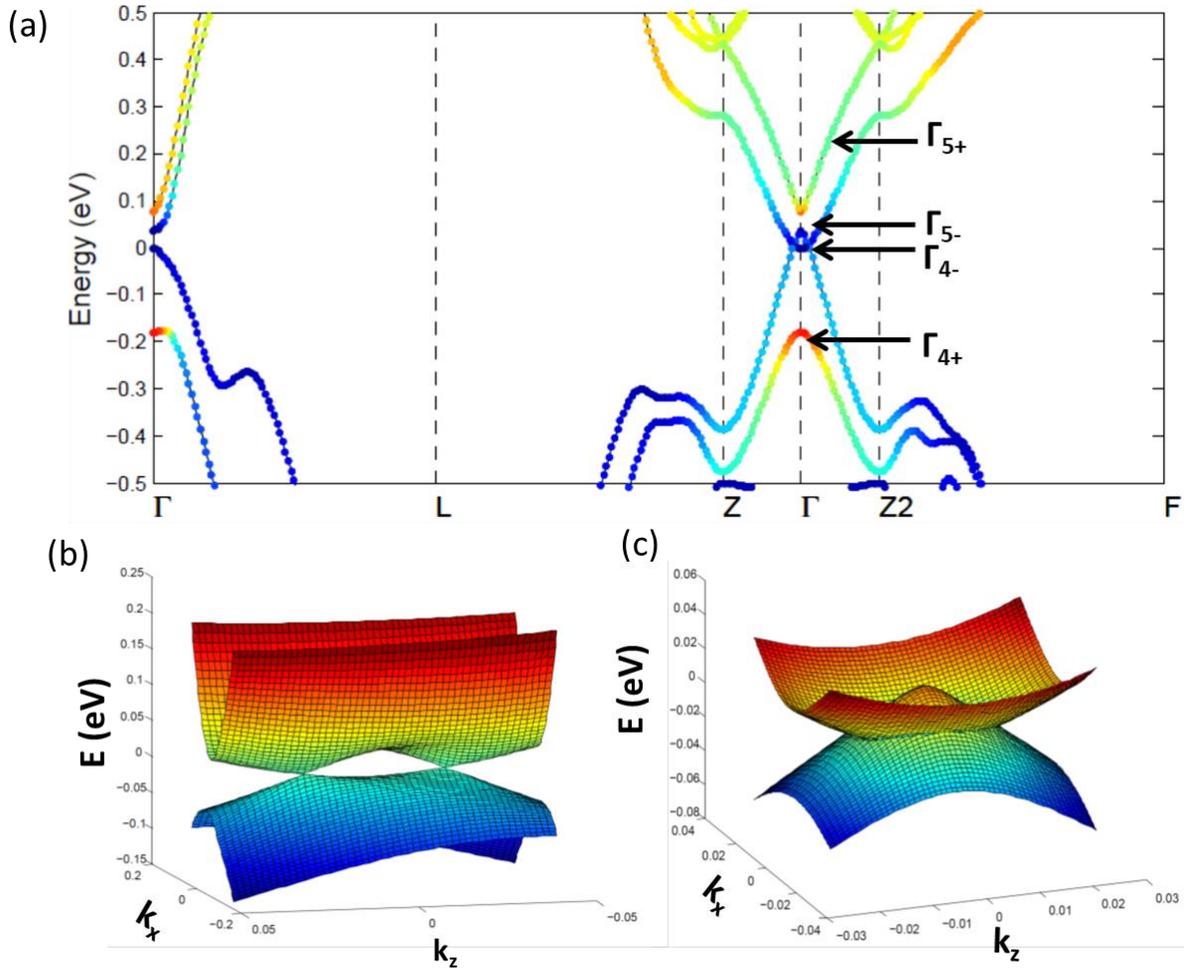

**Fig 3**. (a) The bulk band structure of ferromagnetic $Bi_2MnSe_4$ with spin aligned along $\hat{z}$ direction. Colors represent projection onto Bi-centered Wannier functions. (b) and (c): Energy spectrum of ferromagnetic $Bi_2MnSe_4$ near Γ in the ky-kz plane with Energy along the z axis. Spins along z in (b), spins along x in (c).

Next, we consider another ferromagnetic spin ordering, but with spins in-plane along the (100) direction. The band structure is very similar to the previous case, except near the band crossing, where as shown in Fig. 3b, instead of two Weyl points, we find an elliptical nodal line around the Γ point in the ky, kz plane. This nodal line is due to the fact unlike in the previous case, aligning spins in the (100) direction does not break mirror symmetry in this plane. Even after considering the effects of spin-orbit coupling, the bands in this plane can be classified by their mirror eigenvalues, which results in a nodal line in the presence of band inversion. Unlike the Weyl points, the nodal line is protected by symmetry. We see that the sensitivity of the Fermi surface of this material to the direction of an external field makes it a rich playground for topological behavior, all relating to the band inversion at Γ.

Now, we will discuss the antiferromagnetic (AFM) phase of $Bi_2MnSe_4$, the band structure of which is shown in Fig. 4a. Unlike the ferromagnetic phases, AFM $Bi_2MnSe_4$ has a symmetry operation that combines TRS with a partial translation along the z direction, and therefore allows for a classification by a $Z_2$ invariant for layered antiferromagnetic topological insulators.[76] Due to band inversion at Γ, we find that the AFM phase is a non-trivial antiferromagnetic topological insulator. However, we note that, unlike a typical non-magnetic topological insulator, the (001) surface of AFM BMS will break time-reversal, due to the loss of translation symmetry along z. This change leads to surface properties that differ from a typical $Z_2$ material, as the Dirac cone feature that is characteristic of topological insulators, including $Bi_2PbSe_4$ as shown in Fig. 2b, is protected by TRS, and is absent in this material. This intrinsically broken TRS on the surface makes AFM $Bi_2MnSe_4$ an ideal system to observe some of the more exotic magneto-electric behaviors of topological insulators, that are usually hidden by the metallic surface states. Here we can observe these behaviors without needing to engineer an interface between a topological insulator and a magnetic insulator. For instance, a thick slab of AFM BMS will display the half-integer quantum anomalous Hall effect, where both the top and bottom surfaces will contribute +/- ($e^2/2h$) to the Hall conductivity, depending on the direction of spins at each surface.[19,76] If there are an even number of layers, the total AHC of the slab will be zero, but each surface will separately contribute $|e^2/2h|$. We demonstrate this effect in Fig. 4b, where we use our Wannier tight-binding model to separate the total AHC of a 20 layers AFM BMS slab into layer-by-layer contributions, using the method for calculating the geometric contribution to the AHC.[77] Each surface contributes nearly +/- ($e^2/2h$), although the convergence to bulk-like behavior is rather slow. If, instead, there were an odd number of layers, then there would be a net magnetic moment in the system, and the two surfaces would give equal contributions to the AHC, $e^2/2h + e^2/2h = e^2/h$, making the slab a 2D Chern insulator.

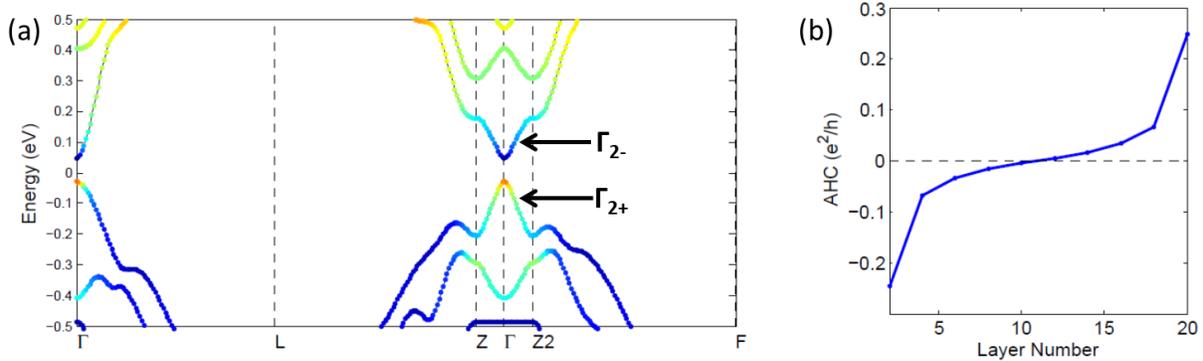

**Fig 4**. (a) The bulk bandstructure of antiferromagnetic $Bi_2MnSe_4$ with spin aligned along $\hat{z}$ direction. Colors represent projection onto Bi-centered Wannier functions. (b) AHC of 20 layers tight-binding slab of antiferromagnetic $Bi_2MnSe_4$, separated into contributions of each double layer, in units of $e^2/h$. Total AHC is zero.

In order to investigate the topological behavior of $Bi_2MnSe_4$ further, we perform first principles slab calculations with a variety of thicknesses and magnetic orders, with (001) surfaces exposed. In the case of a ferromagnetic slab with spins along the z direction, we find that one and two layer thick slabs are topologically trivial, due to confinement increasing the band gap and preventing band inversion. For three layer and thicker slabs, a band inversion occurs at Γ, and the material becomes a Chern insulator with C=-1. In Fig. 5a, we show the band structure of the 3 layers case, which has a band gap of 29 meV. We show other band structures with different thickness of FM and AFM systems in the supplementary materials (Fig S3 and S4). In Fig. 5b, we calculate the edge state of the slab, finding a single spin-polarized edge channel, as expected for a Chern insulator. We also find a Chern insulator in the 4 layers case, with a gap of 58 meV. As the material converges to bulk-like behavior, the gap should eventually close and the AHC will be proportional to the distance between the Weyl points.

For AFM slabs, we find that all slabs from one to five layers are topologically trivial, but with decreasing band gaps from 240 meV for two layers down to 22 meV for five layers, as shown in Fig. 5c. To match the bulk calculations, we expect this trend to continue, and thicker slabs to show a band inversion and a Chern insulating phase, if there are an odd number of layers, as discussed above. Using our Wannier model, we confirm that any perturbation that closes the band gap of the five layers slab, for instance lowering the Bi on-site energies, results in a Chern insulator.

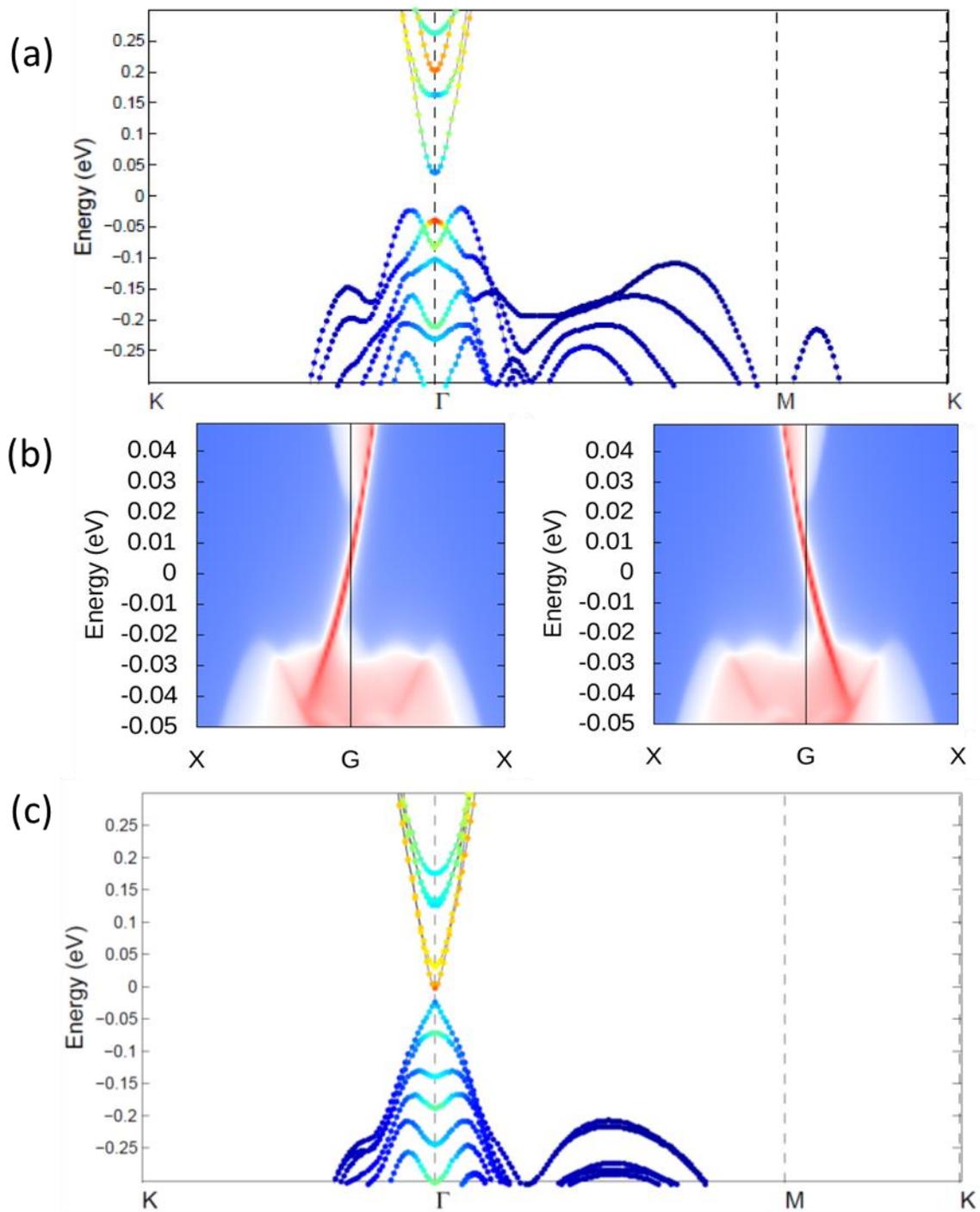

Fig. 5: Band structure of 4SL-$Bi_2MnSe_4$ with FM spins. b) Topologically protected edge states of same system, on left and right edges. c) Band structure of 5SL-$Bi_2MnSe_4$ with AFM spins.

**Conclusion**

We study the topological behavior of $Bi_2MnSe_4$. We find that due to the interplay of band inversion at the Gamma point, strong spin-orbit coupling, and significant spin splitting, $Bi_2MnSe_4$ displays a variety of rare topological phases. These phases include Weyl semimetal, nodal line semimental, AFM topological insulator, and Chern insulator. In thin film form, $Bi_2MnSe_4$ has a non-zero Chern number starting at three layers in the FM phase. We hope that due to its strong magnetic interactions and the fact that it is stochiometric and has significant band gaps, $Bi_2MnSe_4$ will prove a fruitful material for studying many different TRS-broken topological phases that are usually difficult to realize.

The purpose of identifying the computer software in this article is to specify the computational procedure. Such identification does not imply recommendation or endorsement by the National Institute of Standards and Technology.